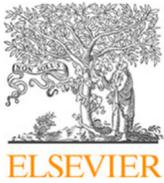
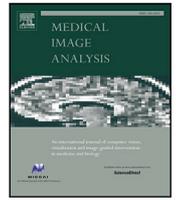

# Error correcting 2D–3D cascaded network for myocardial infarct scar segmentation on late gadolinium enhancement cardiac magnetic resonance images

Matthias Schwab [a],*, Mathias Pamminger [a], Christian Kremser [a], Daniel Obmann [b], Markus Haltmeier [b], Agnes Mayr [a]

[a] *Department of Radiology, Medical University of Innsbruck, Innsbruck, 6020, Tirol, Austria*
[b] *Department of Mathematics, University of Innsbruck, Innsbruck, 6020, Tirol, Austria*

A B S T R A C T

Late gadolinium enhancement (LGE) cardiac magnetic resonance (CMR) imaging is considered the in vivo reference standard for assessing infarct size (IS) and microvascular obstruction (MVO) in ST-elevation myocardial infarction (STEMI) patients. However, the exact quantification of those markers of myocardial infarct severity remains challenging and very time-consuming. As LGE distribution patterns can be quite complex and hard to delineate from the blood pool or epicardial fat, automatic segmentation of LGE CMR images is challenging. In this work, we propose a cascaded framework of two-dimensional and three-dimensional convolutional neural networks (CNNs) which enables to calculate the extent of myocardial infarction in a fully automated way. By artificially generating segmentation errors which are characteristic for 2D CNNs during training of the cascaded framework we are enforcing the detection and correction of 2D segmentation errors and hence improve the segmentation accuracy of the entire method. The proposed method was trained and evaluated on two publicly available datasets. We perform comparative experiments where we show that our framework outperforms state-of-the-art reference methods in segmentation of myocardial infarction. Furthermore, in extensive ablation studies we show the advantages that come with the proposed error correcting cascaded method. The code of this project is publicly available at https://github.com/matthi99/EcorC.git.

## 1. Introduction

Although mortality has been declining over the past decades, ischemic heart disease remains the primary cause of death worldwide, being responsible for an estimated 9.1 million global deaths in 2019 (Nowbar et al., 2019; Vos et al., 2020). After ST-segment elevation myocardial infarction (STEMI), the assessment of infarct size is critical for prognosis of major adverse cardiovascular events (MACE) (Larose et al., 2010) and for clinical decision making prior revascularisation (Kim et al., 2000; Gerber et al., 2012). Furthermore, microvascular obstruction (MVO) was identified as prognostic factor independent of infarct size after reperfused STEMI (de Waha et al., 2017). Cardiac magnetic resonance (CMR) late gadolinium enhancement (LGE) imaging is the reference standard for imaging of ischemic myocardial damage and microvascular obstruction (Kramer et al., 2020). However, in order to obtain these important biomarkers, segmentation of LGE CMR images is necessary, as illustrated in Fig. 1.

Different methods for manual or semi-automated LGE quantification on LGE images have been proposed. Manual LGE segmentation by an expert reader is time consuming and requires particular training with limited reproducibility (Flett et al., 2011). Semi-automated methods for LGE quantification rely on placement of a region of interest (ROI) either in the remote myocardium for the signal threshold versus reference mean (STRM) method, also referred to as the standard deviation (SD) method, or in the infarcted myocardium for the full-width at half-maximum (FWHM) method (McAlindon et al., 2015). Although aiming at reproducibility, ROI placement and size strongly influence signal-intensity thresholds for semi-automated methods (Heiberg et al., 2022). Therefore, fast and reproducible methods for LGE and MVO segmentation after STEMI are desirable.

To circumvent the necessity of manually drawn ROIs, extended intensity based methods have been proposed. Frequently they were based on Otsu thresholding (Tao et al., 2010) or clustering (Detsky et al., 2009). Also fitting myocardial tissue intensities to expected distri-

---

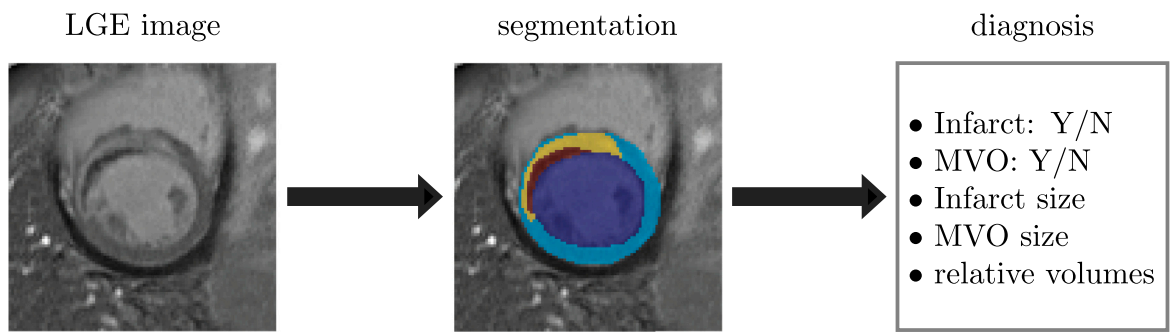

**Fig. 1.** Evaluation of myocardial injury after infarction using LGE CMR. In the short axis images of the left ventricle the tissue gets divided into blood pool (blue) healthy myocardium (cyan), scar tissue (yellow) and MVO (red). Using this segmentation important biomarkers like infarct size or MVO size can be calculated which help to improve clinical decision making as well as predictability for subsequent MACE.

butions using Gaussian mixture models was proposed (Engblom et al., 2016; Liu et al., 2017). In general these intensity based methods have not achieved clinical adoption as these frameworks do not incorporate any spacial context and hence are very sensitive to noise or imaging artifacts.

As in recent years convolutional neural networks (CNNs) have become state-of-the-art in many image segmentation problems, they were also increasingly used in LGE segmentation (Zabihollahy et al., 2018; Moccia et al., 2019; Fahmy et al., 2020). Recently, the topic received even more attention, as two challenges were held in the course of the MICCAI conference in 2020. This included a multi-sequence CMR based myocardial pathology segmentation (MyoPS) challenge (Zhuang and Li, 2020; Li et al., 2023) with the goal of combining multi-sequence CMR data (bSSFP, LGE and T2 CMR) to segment different aspects of ischemic myocardial pathology, including normal myocardium, infarction and edema. Another open-source dataset was made available as part of the EMIDEC challenge (Lalande et al., 2020, 2022), a segmentation contest to compare the performance of automatic methods on the segmentation of the myocardium for LGE CMR exams.

More recently, several research works have been focusing on also including prior knowledge into their segmentation method. This was done by incorporating shape, anatomical, spacial or classification prior information into the segmentation frameworks (Yue et al., 2019; Zhou et al., 2021b; Brahim et al., 2022; Popescu et al., 2022). Furthermore, also cascaded pipelines have proven to get quite popular lately and have shown good results (Zabihollahy et al., 2020; Zhang, 2021; Lustermans et al., 2022).

The U-Net architecture (Ronneberger et al., 2015) and its extensions have become one of the most popular architectures for biomedical image segmentation and are also widely used for LGE segmentation. For segmentation of three-dimensional biomedical data usually the 3D U-Net is the more suitable method, since it is able to extract volumetric information in all spacial directions and hence is usually outperforming its two-dimensional counterpart, which focuses only on intra-slice information. However, the voxel spacing of the 3D volumes in LGE CMR images usually is highly anisotropic. For example in the EMIDEC dataset the median voxel spacing is 1.458 mm $\times$ 1.458 mm $\times$ 10 mm. This means that the resolution in $x$ and $y$ direction is almost 7 times higher than in $z$ direction and hence treating the three axis all equally is far from optimal. Therefore, Zhang (2021) proposed a cascaded pipeline trying to combine the strengths of two and three-dimensional segmentation methods. Recently, 2D–3D cascaded methods have also been proposed in other areas of medical image segmentation, such as glioma segmentation (Cao et al., 2022). However, in our opinion, these methods still suffer from some weaknesses as they do not guarantee perfect synergy between 2D and 3D networks. In experiments we investigated that errors made by the 2D method on a full MR volume are quite characteristic due to its two-dimensional structure but are sometimes not detected by the subsequent 3D network. We therefore claim that it is possible to detect and correct these 2D-characteristic errors with an improved training strategy for the 3D network. If, for example, the 2D segmentation in one single slice deviates very strongly from the neighboring ones, there is a relatively high probability that an error is present in this slice. However, when training a 2D–3D cascaded framework such characteristic errors are quite rare since the performance of the 2D network was optimized on the same training dataset. We therefore propose to include artificially created 2D-characteristic segmentation errors during training of the cascaded framework. This strengthens the ability of the resulting method to detect and correct 2D-segmentation errors which further improves the generalizability properties of the method on unseen data. By constructing characteristic 2D segmentation mistakes we implicitly include prior information into our method. For example, deleting LGE segmentation in one random slice of the 3D volume can be interpreted as implicitly including prior information about the three-dimensional structure of the segmentation objects. In some sense this strategy can also be seen as an additional data augmentation for the 3D network working similar to "random erasing" data augmentation (Zhong et al., 2020), which occludes an arbitrary region of the input image during training and has shown reasonable improvements on object detection and person re-identification tasks.

To summarize, in this work we propose a training framework that is designed to overcome the disadvantage of error propagation in cascaded methods. The main contributions of this work are as follows:

- We present a novel perturbation module for efficient training of cascaded 2D–3D segmentation pipelines, which improves the generalization properties of the final method to unseen data.
- We test our method on two publicly available datasets and show that it outperforms current state-of-the-art methods especially regarding segmentation of myocardial infarction.
- We show the advantages that come when training a cascaded framework with the introduced perturbation module in various ablation experiments.

The rest of the paper is organized as follows. Section 2 describes the proposed error-correcting segmentation framework for automated scar pattern detection in LGE CMR volumes. In Section 3, the experimental results achieved on the used datasets are presented. Section 4 discusses results, limitations and potential directions for future works. Finally, Section 5 gives a short summary of the presented research.

## 2. Material and methods

### 2.1. Datasets

For this work two publicly available LGE CMR datasets were used:

The CMR database from the **EMIDEC** challenge consists of 150 clinical exams containing healthy patients (1/3) and patients with myocardial infarction (2/3). Each exam includes a series of LGE CMR images in short axis orientation covering the left ventricle from the base to the apex. The CMR acquisition was performed on 1.5T and 3T Siemens MRI scanners at the University Hospital of Dijon (France). All measurements were performed ECG-gated and were carried out 10 min after gadolinium based contrast agent injection using a T1-weighted phase sensitive inversion recovery (PSIR) sequence (TR = 3.5 ms, TE = 1.42 ms, TI = 400 ms, flip angle = 20). The data were divided into training ($n = 100$) and testing ($n = 50$) datasets where ground truth segmentation masks were only provided for patients in the training set. The manual delineations were outlined by two experts (a cardiologist with 10 years of experience and a biophysicist with 20 years of experience) and include the endocardial and epicardial borders, the infarcted areas as well as the areas of MVO, if present. To summarize, the ground truth annotations divide the images into five different classes: blood pool, healthy myocardium, myocardial scar, MVO and background. Exact details on the EMIDEC dataset can be found in Lalande et al. (2020).

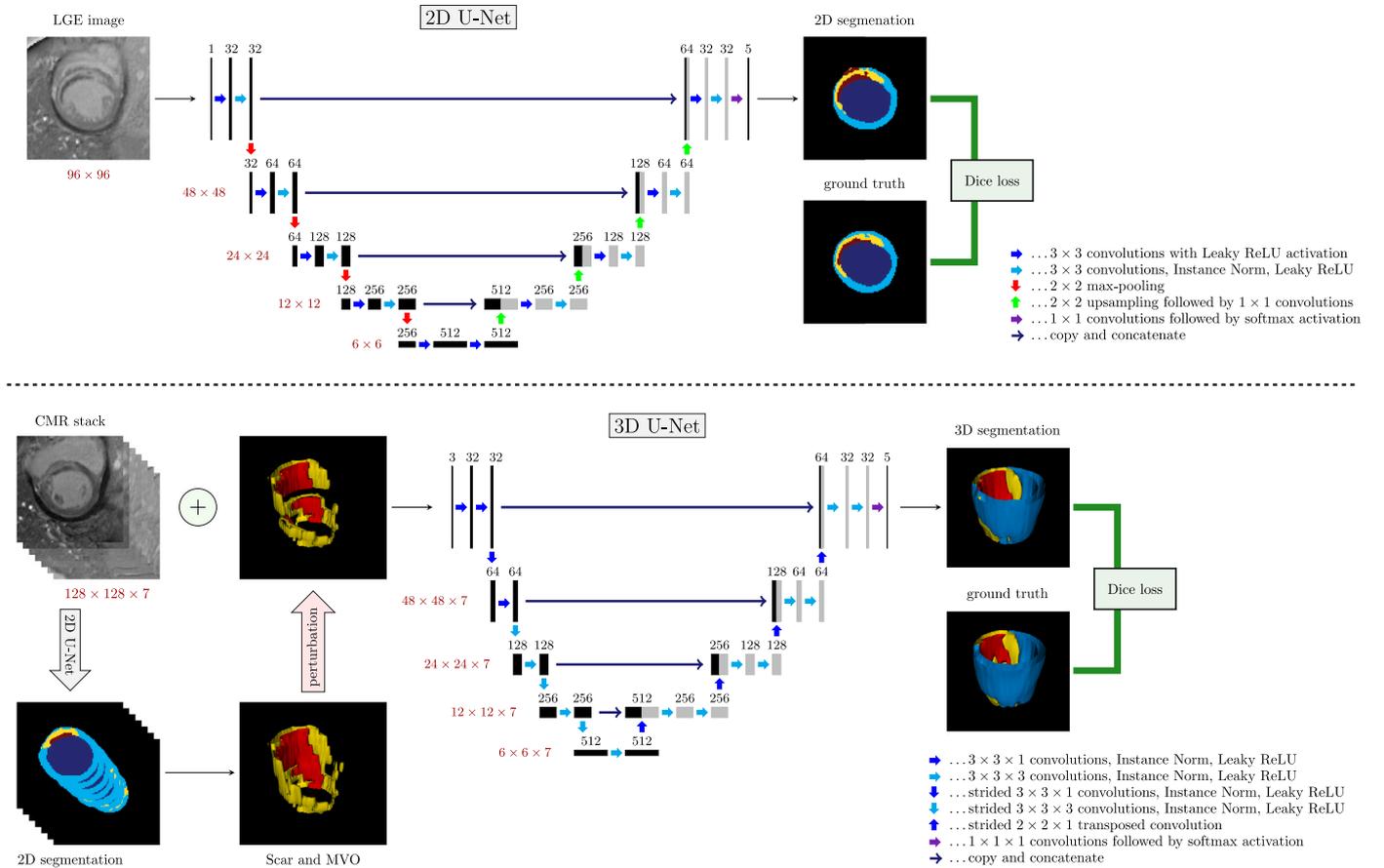

**Fig. 2.** The proposed segmentation pipeline. Firstly a 2D U-Net is trained on the individual images of the training dataset. After that a 3D U-Net is trained to refine the segmentation and return the final three-dimensional segmentation volume for the left ventricle. Input for the 3D CNN are the LGE CMR volumes as well as the possibly perturbed segmentation masks for scar and MVO provided by the pre-trained 2D CNN.

The dataset provided by the **MyoPS** challenge consists of 45 paired three-sequence CMR images. These include bSSFP, LGE and T2 sequences in short axis orientation acquired from a 1.5T Philips Achieva scanner at the Shanghai Renji hospital. The dataset only contains male patients with myocardial infarction and was divided into a training ($n = 25$) and a test ($n = 20$) dataset. All the sequences have been preprocessed using a multivariate mixture model method (Zhuang, 2019) to align the three sequences and to resample them into the same spatial resolution. Gold standard segmentations of the sequences were done by post-graduate students either in biomedical engineering or medical imaging. Further, all manual segmentations were checked by three experts in cardiac anatomy. All together the manual annotation include five different labels: left and right ventricular blood pools, left ventricular myocardium, scar and edema. As our method was not developed for multi sequence segmentation we only use the LGE images for our experiments. Since the edema is only visible in the T2 sequences we neglect this label and only divide into left ventricular blood pool, healthy myocardium and scar.

### 2.2. Preprocessing

Since both the EMIDEC and the MyoPS dataset exclusively contain CMR slices in which myocardium is present no extensive preprocessing pipeline is necessary. To get a bigger proportion of foreground pixels per sample the original input images are cropped around the center of the left ventricle. After that all the images are normalized independently to have zero mean and a standard derivation of one. For the three dimensional stacks input sizes of $96 \times 96 \times 7$ and $320 \times 320 \times 5$ are chosen for the EMIDEC and the MyoPS dataset, respectively. For volumes containing more slices than the chosen input size this is achieved by randomly selecting subvolumes contiguous slices during training. For examinations containing less slices the volume is resized using nearest neighbor interpolation with respect to the $z$- axis.

### 2.3. Network architecture

The main architecture of our method consists of a cascade of 2D and 3D U-Nets. The framework can therefore be divided into two main steps.

1. Segmentation on the individual slices of the CMR volume by a two-dimensional CNN.
2. Refinement of the 2D segmentation masks with the help of a three-dimensional CNN.

For both networks we use basic U-Net architectures (Ronneberger et al., 2015) with kernel sizes of $3 \times 3$ and $3 \times 3 \times 3$, respectively. After each convolutional block we apply instance normalization followed by a leaky rectified linear unit (ReLU) activation function. For the 2D U-Net downsampling is achieved by max-pooling and upsampling by bilinear interpolation. The 3D U-Net uses strided convolutions, following (Szegedy et al., 2016), for downsampling as well as strided convolution transposed for upsampling. Since the image size is already very small in the third dimension no downsampling or upsampling is performed regarding this dimension. To prevent the 3D method from merely copying the segmentation masks provided by the 2D network, we only add the 2D segmentation masks for infarction and MVO as additional input, leaving segmentation of healthy myocardium and blood pool to the 3D network alone. The overall configuration of our framework is illustrated in Fig. 2.

## 2.4. Postprocessing

To be able to objectively assess the influence of different methods on the final segmentation result we refrain from any postprocessing steps. We simply take the argmax over the different segmentation channels as our final class prediction. After that we zero pad the predicted segmentation masks to bring them back to the original field-of-view.

## 2.5. Data augmentation

To provide a stream of constantly changing training examples we apply various data augmentation techniques during training. This includes intensity based transformations such as Gaussian blurring, gamma correction, additive white noise, changing contrast and brightness as well as simulating images with lower resolution. In addition to this we also use various geometric transformations including translation, flipping, elastic deformation and scaling to further improve the data variety during training.

## 2.6. Perturbation module

One of the main advantages of the cascaded pipeline is that potential segmentation errors done in the first stage can be corrected afterwards. However, one big issue arising during training is that the coarse segmentation masks provided as additional input are already quite precise, since 2D-net was optimized on the same training dataset. Therefore, we introduce a novel perturbation module which ensures that segmentation errors made by the first network due to missing information about neighboring slices are detected and improved by the subsequent network.

The idea is to artificially create different 2D-characteristic segmentation errors during training in order to then allow the 3D network to learn to correct them. This includes:

- Slightly increasing some data augmentation parameters compared to the 2D augmentation pipeline. More precisely this means increasing the ranges from which contrast, brightness and gamma augmentations get chosen as well as producing images of even lower resolution.
- Randomly deleting the 2D segmentations for a specific class (myocardial scar or MVO or both) on single slices (see Fig. 3 second and third row).

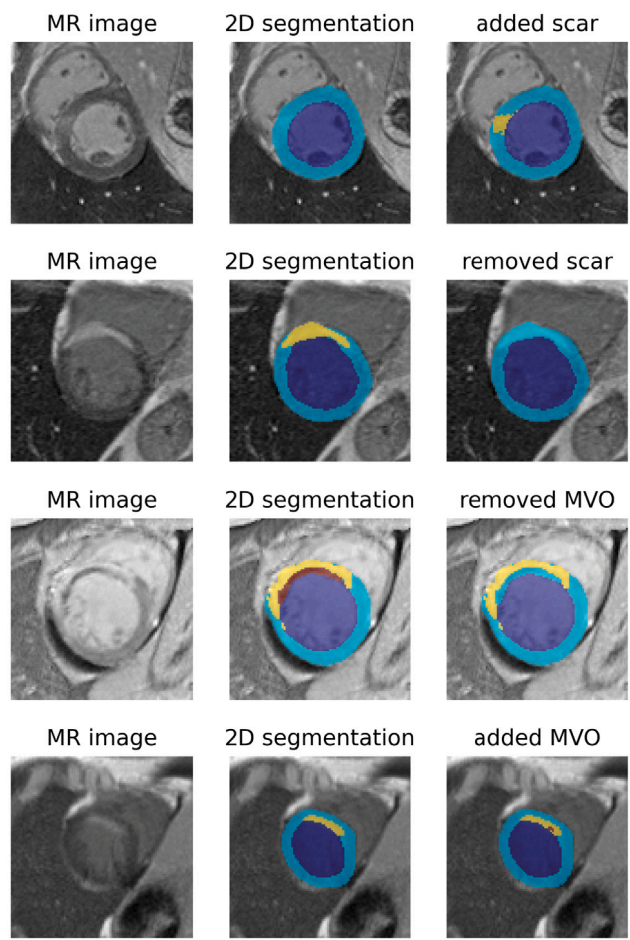

Fig. 3. Examples for the artificial modifications of the 2D segmentation masks. The first row shows an example where the 2D segmentation correctly segmented a healthy patient, however, an artificial scar was created by the perturbation module. Row two shows a contrary example were scar tissue was at first correctly identified but was removed afterwards. Rows three and four show examples for removed resp. added MVO segmentation.

- Adding wrong annotations for infarction to single slices. If this is triggered for a sample a random slice of the 3D-volume is chosen and the 85th percentile of the pixel values of the myocardium in this slice is calculated using the ground truth segmentation mask. After thresholding regarding this percentile we add the biggest connected component as scar to this specific slice (see Fig. 3 first row).
- Adding wrong MVO annotations to single slices. Since MVO always occurs inside an infarcted region, we are generating fake MVO annotations by randomly choosing a pixel which was classified by the 2D framework as scar and change the label of this pixel as well as a few random neighboring pixels to MVO (see Fig. 3 fourth row).
- Setting the whole 2D-segmentation mask to zero for some random samples.

During training of the cascaded pipeline at most one of these operations is chosen randomly with small probabilities. For our experiments we chose probabilities of 10% for deleting single classes, setting the whole 2D segmentation mask to zero as well as for adding wrong infarction annotations. Incorrect MVO annotations have been added with a probability of 2%. As there are no MVO labels present in the MyoPS dataset, we slightly adapt our perturbation module only performing perturbations on the myocardial infarction class. Since most of the perturbations are performed only in single slices of the whole MR volume, the three-dimensional system is specially trained to detect errors that are characteristic for missing inter-slice information.

**Table 1**
Results of the five-fold cross-validation for the proposed method on the EMIDEC dataset.

| Targets | Metrics | Fold 1 | Fold 2 | Fold 3 | Fold 4 | Fold 5 | Mean | SD |
|---|---|---|---|---|---|---|---|---|
| Myocardium | DSC (%) | 87.24 | 86.20 | 87.00 | 85.12 | 85.11 | 86.13 | 0.9 |
|  | AVD (mm$^3$) | 8289 | 8761 | 6920 | 8037 | 12105 | 8823 | 1749 |
|  | HAUS (mm) | 13.95 | 12.78 | 11.39 | 13.50 | 17.46 | 13.82 | 2.02 |
| Infarction | DSC (%) | 76.54 | 70.64 | 79.75 | 77.70 | 75.54 | 76.03 | 3.04 |
|  | AVD (mm$^3$) | 2904 | 5391 | 2054 | 3853 | 5847 | 4010 | 1439 |
|  | AVDR (%) | 2.45 | 3.83 | 1.76 | 3.23 | 4.49 | 3.15 | 0.97 |
| MVO | DSC (%) | 69.33 | 76.27 | 83.66 | 78.33 | 57.72 | 73.06 | 8.94 |
|  | AVD (mm$^3$) | 759 | 1580 | 877 | 356 | 953 | 905 | 395 |
|  | AVDR (%) | 0.64 | 1.03 | 0.63 | 0.35 | 0.70 | 0.67 | 0.22 |

By introducing this kind of perturbations during training we enforce the 3D network to be nearly invariant under erroneous 2D segmentations in single slices. To illustrate this let us denote $(x_i)_{i=1}^M$ as the three dimensional input stack of the CMR, where $M$ is the number of 2D slices for the specific examination. Further we denote the CNNs as $\phi_{2D}$ and $\phi_{3D}$, respectively. Now let $\delta_k$ be a perturbation operator, which is acting on the $k$th slice of a segmentation mask. By training our cascaded framework on disturbed as well as on clean segmentation masks we guide the three dimensional framework to have the property

$$\phi_{3D}\left((x_i)_{i=1}^M, \delta_k\left((\phi_{2D}(x_i))_{i=1}^M\right)\right) \simeq \phi_{3D}\left((x_i)_{i=1}^M, (\phi_{2D}(x_i))_{i=1}^M\right). \quad (1)$$

Interpreting $\delta_k$ as a 2D segmentation error happening in the $k$th slice of the CMR volume Eq. (1) shows the capability of the cascaded framework for correcting segmentation errors on single slices.

### 2.7. Loss

As loss function a variation of the Dice loss (Eq. (2)) as proposed in Milletari et al. (2016) was used. As this loss is equivalent to the negative Dice coefficient for binary images, its main advantage is that it is very useful for establishing the right balance between foreground and background voxels. This enables us to get rid of assigning weights to samples of different classes. Therefore, for all the foreground classes (blood pool, healthy myocardium, scar, MVO) during training we minimize

$$L_{DCE} = -\frac{2\sum_{i \in I} p_i g_i}{\sum_{i \in I} p_i^2 + \sum_{i \in I} g_i^2}, \quad (2)$$

where $I$ is the set of pixels, $p \in \mathbb{R}^I$ is the networks segmentation prediction and $g \in \{0,1\}^I$ is the corresponding ground truth mask. To alleviate the common problem of "vanishing gradients" we use deep supervision (Lee et al., 2015) additionally providing direct supervision to some hidden layers instead of only supervising the output layer.

The detailed training procedure looks as follows. At first we trained the two-dimensional U-Net with a batch size of 32 over 750 epochs. Then the 3D cascade framework was trained for 750 epochs using a batch size of 4. When training the 3D cascade, the perturbation module was switched on after 100 training epochs to further enforce the detection and improvement of 2D segmentation errors of the final method. For minimization of the loss, stochastic gradient descent with Nesterov momentum ($\mu = 0.99$) and initial learning rate of 0.005 (2D) and 0.01 (3D), respectively was used. All the training of the framework was performed on a NVIDIA A40 GPU using the Pytorch deep learning library. Detailed descriptions of the implementation and exact parameter configurations can be found in *Supplementary Materials*.

### 2.8. Reference methods

We compare our method against four other approaches which we briefly describe in this section.

- **nnU-Net** (Isensee et al., 2021) is a deep learning based segmentation framework for biomedical images. It is a self-configuring method which has proven to be state-of-the-art in biomedical image segmentation on a wide variety of different datasets and is very commonly used as a baseline method. At MICCAI 2020, for example, 9 out of 10 challenge winners built their methods on top of nnU-Net. Since the method can handle different image dimensions, we compare our method to both 2D and 3D nnU-Net.
- To also consider network architectures other than CNNs, we compare our method with **nnFormer**, a 3D transformer network for volumetric medical image segmentation (Zhou et al., 2021a). As the architecture of nnFormer maintains a U-shape it exploits the combination of interleaved convolution and self-attention operations. Further, the traditional concatenation and summation operations in the skip connections of the U-Net are replaced by an attention mechanism called skip attention.
- Since we use the dataset of the EMIDEC Challenge in this paper, we also compare our method with that of the **winner of the challenge** (Zhang, 2021). Similar to us, this method uses a cascaded pipeline for image segmentation, where simply the full 2D segmentations are added as additional channels to a 3D network without any special attention to whether this information is used or not for the final prediction.
- Finally, to position our method in the current research landscape, we compare our framework with a very **recently published** method (Rahman and Marculescu, 2023). In this work, the authors introduced a Cascaded Attention Decoder, a novel attention-based decoder that showed promising improvements for medical image segmentation tasks using vision transformers.

### 2.9. Evaluation metrics

To quantify the segmentation accuracy of our method we stick to the metrics proposed by the organizers of the EMIDEC challenge, which incorporates both clinical and geometrical metrics. For the evaluation of the myocardial tissue this includes Dice coefficient (DSC), Hausdorff-distance (HAUS) in mm as well as absolute volume difference (AVD) in mm$^3$. The Dice coefficient (Eq. (3)) is a measure of the spacial overlap between two sets $P$ and $G$. It is defined as

$$DSC = \frac{2|P \cap G|}{|P| + |G|}, \quad (3)$$

where $|\cdot|$ denotes the cardinality of a set. In our case $P$ presents the final segmentation prediction of our method and $G$ the manual gold standard. The range of the Dice coefficient resides between 0 (no overlap at all) and 1 (perfect match). The Hausdorff-distance (Eq. (4)) also is a common metric evaluating the degree of mismatch between

**Table 2**
Results of the five-fold cross-validation for the proposed method on the MyoPS dataset.

| Targets | Metrics | Fold 1 | Fold 2 | Fold 3 | Fold 4 | Fold 5 | Mean | SD |
|---|---|---|---|---|---|---|---|---|
| Myocardium | DSC (%) | 85.55 | 83.04 | 81.33 | 81.04 | 83.24 | 82.82 | 1.78 |
|  | AVD (mm$^3$) | 9065 | 21 874 | 14 682 | 18 054 | 12 225 | 15 180 | 4986 |
|  | HAUS (mm) | 13.07 | 14.91 | 15.08 | 17.29 | 11.89 | 14.45 | 2.07 |
| Infarction | DSC (%) | 68.41 | 62.44 | 59.17 | 67.36 | 52.62 | 62.00 | 6.44 |
|  | AVD (mm$^3$) | 7538 | 9944 | 11 131 | 10 118 | 11 569 | 10 060 | 1565 |
|  | AVDR (%) | 4.96 | 8.53 | 6.99 | 7.27 | 9.28 | 7.41 | 1.66 |

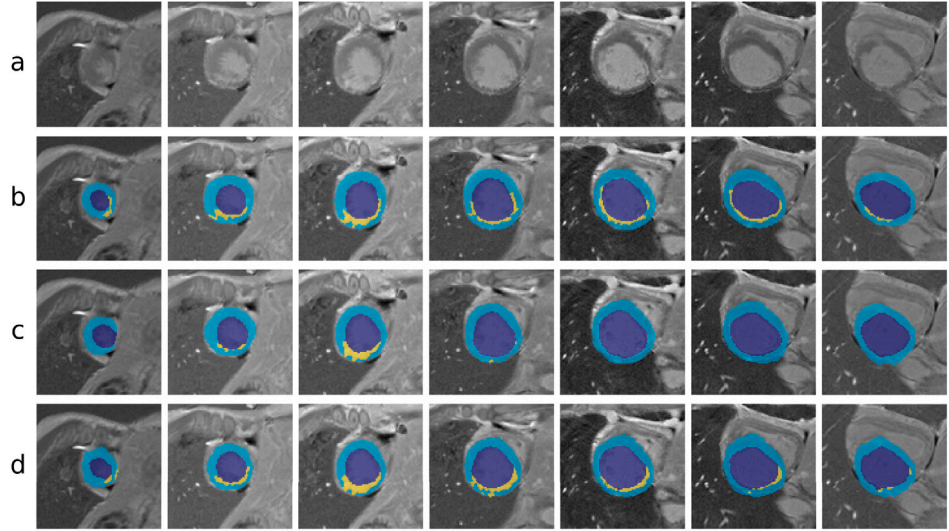

**Fig. 4.** Example of an LGE CMR stack (a) with according manual segmentation (b). Scar tissue is missed by the 2D method (c) in some slices (DCE = 36.57%). The cascaded method (d) is able to improve the segmentation performance (DCE = 60.59%), correctly identifying additional scar tissue in neighboring slices.

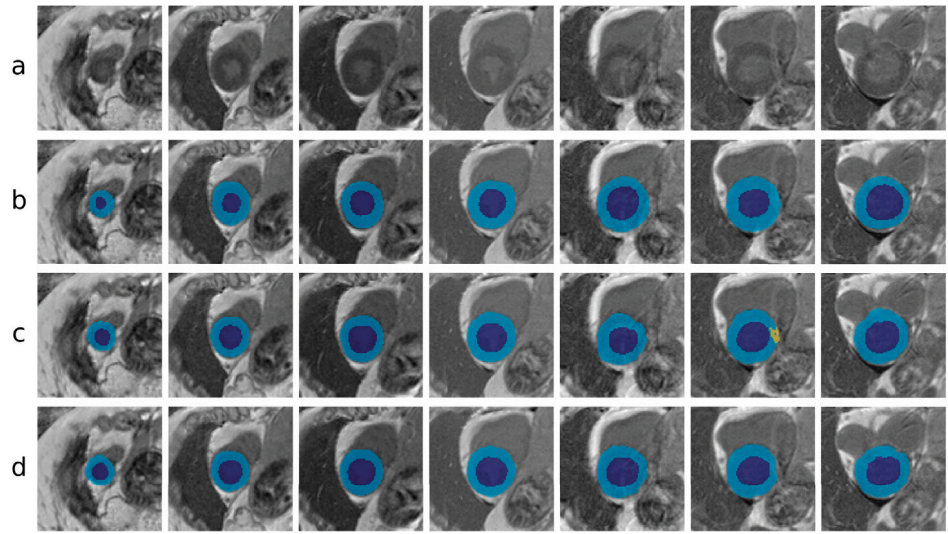

**Fig. 5.** Example of a healthy patient. The wrongly identified scar tissue gets recognized and removed by the cascaded pipeline. (a) LGE MR image, (b) ground truth, (c) 2D segmentation, (d) 3D cascade segmentation.

two segmentation boundaries by calculating Euclidean distances. It is defined as

$$\mathrm{HAUS} = \max\{\max_{p \in P} d(p, G), \max_{g \in G} d(g, P)\}, \qquad (4)$$

where $d(a, B) = \min_{b \in B} \|a - b\|$ quantifies the minimal euclidean distance from a point $a \in A$ to set $B$. Further, the absolute volume difference (Eq. (5)) calculates the amount of difference between the volume $V_A$ obtained by the proposed method and the volume $V_M$ calculated from the manual segmentations

$$\mathrm{AVD} = |V_A - V_M|. \qquad (5)$$

For quantification of infarction and MVO volumes additionally the absolute volume difference rate (AVDR) according to the volume of the myocardium, which we denote by $V_{\mathrm{MYO}}$, is calculated (Eq. (6)). To ensure consistency with the results of the reference methods, the evaluation scores were calculated using the official evaluation code of

Table 3
A comparison of evaluation scores of different methods on a five-fold cross-validation on the EMIDEC dataset.

| Targets | Metrics | 2D nnU-Net | 3D nnU-Net | Zhou et al. (2021a) | Zhang (2021) | Rahman and Marculescu (2023) | Ours |
|---|---|---|---|---|---|---|---|
| Myocardium | DSC (%) | 85.50 ± 3.70* | **87.21 ± 3.63** | 85.44 ± 4.14* | 86.28 ± 4.36* | 86.44 ± 4.15* | 86.13 ± 4.04* |
| | AVD (mm$^3$) | 7564 ± 6949 | 6879 ± 5718 | 6681 ± 6321 | **6618 ± 5537** | 7864 ± 6674* | 8823 ± 6494* |
| | HAUS (mm) | 27.96 ± 79.66 | 14.16 ± 8.38 | 25.98 ± 87.22 | 15.01 ± 10.19 | **13.43 ± 6.09** | 13.82 ± 6.92 |
| Infarction | DSC (%) | 50.85 ± 33.12* | 72.95 ± 24.07 | 62.02 ± 30.27* | 73.40 ± 23.36* | 59.13 ± 31.97* | **76.03 ± 20.49** |
| | AVD (mm$^3$) | 5610 ± 8397* | 4469 ± 6416 | **3861 ± 5751** | 4239 ± 5992 | 4682 ± 6664* | 4010 ± 6161 |
| | AVDR (%) | 4.53 ± 5.56* | 3.57 ± 4.33 | **3.15 ± 4.25** | 3.28 ± 3.94 | 3.79 ± 4.56* | 3.15 ± 4.35 |
| MVO | DSC (%) | 67.35 ± 40.50* | 71.36 ± 38.91 | 68.06 ± 40.61* | 72.58 ± 37.46 | 64.51 ± 41.85* | **73.06 ± 37.05** |
| | AVD (mm$^3$) | 1058 ± 3253 | 969 ± 2969 | 1024 ± 2652 | 939 ± 2783 | 1175 ± 3321 | **905 ± 2584** |
| | AVDR (%) | 0.78 ± 1.85 | 0.71 ± 1.69 | 0.75 ± 1.65 | 0.69 ± 1.57 | 0.86 ± 1.92 | **0.67 ± 1.48** |

* Indicates the method obtained statistically poorer results ($p < 0.05$) compared to the best performance (marked in bold) in terms of the corresponding metrics.

Table 4
A comparison of evaluation scores of different methods on a five-fold cross-validation on the MyoPS training dataset.

| Targets | Metrics | 2D nnU-Net | 3D nnU-Net | Zhou et al. (2021a) | Zhang (2021) | Rahman and Marculescu (2023) | Ours |
|---|---|---|---|---|---|---|---|
| Myocardium | DSC (%) | 84.28 ± 3.93 | **84.58 ± 3.95** | 77.41 ± 17.61* | 84.27 ± 4.23 | 81.36 ± 5.827* | 82.82 ± 3.94* |
| | AVD (mm$^3$) | 14 085 ± 11 309 | **13 313 ± 9861** | 24 487 ± 32 514* | 14 098 ± 10 955 | 15 375 ± 15 259 | 15 179 ± 9566 |
| | HAUS (mm) | **11.21 ± 6.60** | 14.34 ± 18.95 | 12.72 ± 6.82 | 11.48 ± 6.59 | 13.43 ± 13.26 | 14.45 ± 10.56 |
| Infarction | DSC (%) | 59.39 ± 19.58 | 58.74 ± 19.56 | 46.54 ± 22.91* | 59.15 ± 19.24 | 55.58 ± 19.15* | **62.00 ± 15.48** |
| | AVD (mm$^3$) | 10 492 ± 10 735 | 10 501 ± 10 874 | 11 358 ± 11 654 | **9975 ± 10 779** | 10 205 ± 9473 | 10 060 ± 9763 |
| | AVDR (%) | 6.11 ± 5.40 | 6.35 ± 4.68 | 6.66 ± 5.64 | **5.85 ± 5.44** | 5.95 ± 5.21 | 7.41 ± 5.28 |

* Indicates the method obtained statistically poorer results ($p < 0.05$) compared to the best performance (marked in bold) in terms of the corresponding metrics.

the EMIDEC challenge (https://github.com/EMIDEC-Challenge/Evaluation-metrics/).

$$\text{AVDR} = \frac{\text{AVD}}{V_{\text{MYO}}}. \quad (6)$$

## 3. Experiments and results

### 3.1. Five-fold cross validation

To test the performance of our method, we split the 100 exams of the EMIDEC dataset into five random subsets and performed a five-fold cross validation. So for each fold we have 80 exams available for training and 20 exams for evaluation. The results that our method achieved on the different folds are displayed in Table 1. Taking the mean over all folds we achieved Dice coefficients of 86% for the entire myocardium, 76.03% for infarcted tissue and 73.06% for MVO. The mean Hausdorff distance between predicted and ground truth myocardium was 13.82 mm$^3$ and the average AVDR with respect to myocardial volume was 3.15% for infarction and 0.67% for MVO, respectively. Also the standard deviations (SD) between the results on the different folds are displayed in Table 1. The same five-fold cross-validation we performed on the 25 training exams of the MyoPS dataset where we find slightly worse segmentation accuracies regarding all three evaluation metrics, as shown in Table 2. For example, for myocardium and scar segmentation we achieved mean Dice coefficients of 82% and 62%, respectively.

### 3.2. Comparison studies

We further compared the proposed method to the reference methods described in Section 2.8. The mean values and standard deviations considering the five-fold cross-validation regarding the different metrics are displayed in Table 3 (EMIDEC) and Table 4 (MyoPS). Considering the Dice coefficients of myocardial infarction segmentation our methods outperforms all reference methods on both datasets. On the EMIDEC dataset the achieved Dice scores of our method are significantly better ($p < 0.05$) compared to all reference methods except the 3D nnU-Net ($p = 0.052$). Also for MVO segmentation our method reports the best Dice scores, however only outperforming three (2D nnU-Net, Zhou et al., 2021a; Rahman and Marculescu, 2023) out of the five reference methods significantly. Although our method has been

Table 5
Comparing Dice coefficients for scar segmentation on the test dataset of the MyoPs challenge.

| Methods | DSC (%) |
|---|---|
| 2D nnU-Net | 62.95 ± 24.89 |
| 3D nnU-Net | 64.20 ± 22.08 |
| Zhou et al. (2021a) | 55.11 ± 27.21* |
| Zhang (2021) | 62.04 ± 25.34 |
| Rahman and Marculescu (2023) | 60.20 ± 28.18* |
| Ours | **66.21 ± 21.33** |

* Indicates the method obtained statistically poorer results ($p < 0.05$) compared to the best performance, which is marked in bold.

optimized to segment infarcted tissue as accurately as possible, we also compare the performance for segmentation of myocardium. For this target, it turns out that on both datasets 3D nnU-Net is able to achieve the best Dice scores even outperforming our method significantly. As an evaluation kit was released for the MyoPS-challenge recently, we were also able to report Dice coefficients for scar segmentation on the test dataset of the challenge. Here our method yielded a more than 2% higher mean Dice value compared to all other reference methods, as displayed in Table 5.

### 3.3. Ablation studies

To investigate the impact of the individual steps of the whole pipeline we also performed ablation studies. We investigated the influence of the perturbation module by training exactly the same cascaded pipeline with the only difference that during training the 2D segmentation masks of scar and potentially MVO were passed unchanged to the 3D U-Net. We refer to this method as "vanilla cascade". It is worth to mention that both networks trained with and without perturbation module were trained using the 2D segmentation outputs of our 2D U-Net and never saw any segmentation outputs of the reference methods during training. We then tested to what extent the two methods were able to improve the segmentation accuracies of the different reference methods. The detailed results for segmentation of myocardial infarction on the MyoPs test dataset are displayed in Table 7, while scores achieved on the EMIDEC dataset can be found in Table 6. Regarding the segmentation of myocardial infarctions in both datasets, the network trained with our perturbation module was able

Table 6
Ablation study on the EMIDEC dataset. A comparison study of Dice scores regarding segmentation of myocardial infarction and MVO when trying to improve accuracy using a cascaded framework.

| Target | Methods | 2D Net Ours | 2D nnU-Net | 3D nnU-Net | Zhou et al. (2021a) | Zhang (2021) | Rahman and Marculescu (2023) |
|---|---|---|---|---|---|---|---|
| Infarction | Original | 68.81 ± 26.95* | 50.85 ± 33.12* | 72.95 ± 24.07* | 62.02 ± 30.27* | 73.4 ± 23.36* | 59.13 ± 31.97* |
|  | Vanilla-cascade | 72.25 ± 25.11* | 62.01 ± 31.83* | 74.69 ± 23.55 | 65.70 ± 29.26* | 74.66 ± 22.73 | 65.34 ± 30.62* |
|  | Ours | **76.03 ± 20.49** | **71.17 ± 26.62** | **76.04 ± 21.66** | **69.99 ± 26.78** | **76.14 ± 20.42** | **74.25 ± 24.03** |
| MVO | Original | 71.71 ± 38.33* | 67.35 ± 40.50* | 71.36 ± 38.91* | 68.06 ± 40.61* | 72.58 ± 37.46* | 64.51 ± 41.85* |
|  | Vanilla-cascade | 72.83 ± 37.71 | **69.65 ± 39.41** | **72.69 ± 37.91** | 69.14 ± 40.05 | 72.4 ± 37.46 | **65.67 ± 41.14** |
|  | Ours | **73.06 ± 37.05** | 69.55 ± 39.50 | 72.00 ± 38.58 | **69.28 ± 39.69** | **73.99 ± 36.33** | 65.01 ± 41.81 |

* Indicates the method obtained statistically poorer results ($p < 0.05$) compared to the best performance, which is marked in bold.

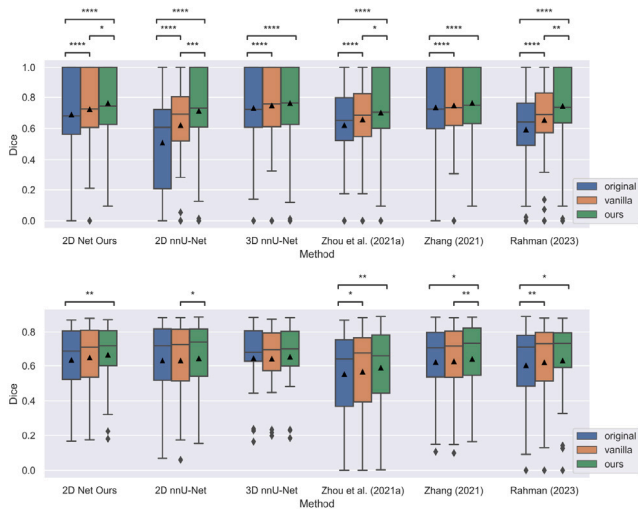

**Fig. 6.** Box plots of Dice coefficients for scar segmentation on the EMIDEC (top) and MyoPS (bottom) datasets. Asterisks report statistical significance values ($p < 0.05^*$, $p < 0.01^{**}$, $p < 0.001^{***}$, $p < 0.0001^{****}$).

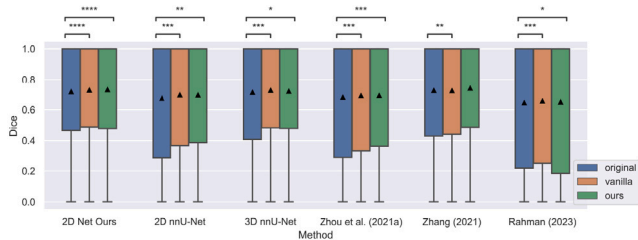

**Fig. 7.** Box plots of Dice coefficients for MVO segmentation on the EMIDEC dataset. Asterisks report statistical significance values ($p < 0.05^*$, $p < 0.01^{**}$, $p < 0.001^{***}$, $p < 0.0001^{****}$).

to improve segmentation accuracy the most for every single reference method. Box plots for the Dice coefficients regarding the infarction class are shown in Fig. 6. Also statistical significance information is displayed. Although the gaps between the scores are smaller for the MyoPS dataset, the vanilla cascade framework almost always improved segmentation performance for myocardial infarction. Nevertheless, the cascaded framework trained with our perturbation module consistently outperformed the vanilla cascade framework. For MVO segmentation the perturbation module did not have that big of an impact as our framework outperformed vanilla cascade only in three out of the six cases. As displayed in Table 6 and Fig. 7 both cascaded methods were able to improve MVO segmentation significantly. However, no significant difference between the two cascaded methods could be found.

## 4. Discussion

In this work an error correcting 2D–3D cascade framework for myocardial scar segmentation on LGE CMR images was introduced. Using both 2D and 3D segmentation methods this approach is able to optimally combine their strengths. By at first training on two dimensional images the advantage that there are much more 2D images than 3D volumes available for training is used. After that, the three-dimensional CNN exploits both the relationships between the slices as well as preceding 2D segmentations. By introducing various perturbations to the 2D segmentation masks during training we further guide the predictions of the final framework to be invariant under various 2D segmentation errors. The proposed pipeline was trained and tested using images and manual segmentations of two publicly available datasets.

Our method outperformed state-of-the-art methods in both scar and MVO segmentation. Especially in segmentation of myocardial infarction our method outperformed other approaches significantly. Compared with the best values of the reference methods we reached an improvement of Dice scores by around 2 to 4 percent for segmentation of myocardial infarction. Also in MVO segmentation our method performed superior compared to all the reference methods, although the differences did not reach the level of significance. To summarize, on the EMIDEC dataset our method reached mean Dice scores of 76.03% for scar segmentation and 73.06% for MVO segmentation. This Dice scores are higher than scores reported for inter-observer studies and very similar to intra-observer scores. For example, on a separate test dataset consisting of 34 examinations (Lalande et al., 2020) report intra-observer Dice coefficients of 76% and inter-observer Dice coefficients of 69% for myocardial infarction segmentation. Although the Dice overlap of 66.21% was lower for myocardial infarction on the MyoPs test data our method clearly also surpassed the reported (Li et al., 2023) inter-observer variations of 56.9% on this dataset. This shows that our automatic segmentation is of comparable quality to manual segmentation.

A possible disadvantage of cascaded methods is that errors made in the first steps could get propagated through the subsequent steps. This could also be a problem in the cascade of two and three-dimensional segmentations, since it is also conceivable that structures that were drawn incorrectly by the 2D network are adopted almost one-to-one by the subsequent 3D method. However, our proposed perturbation module that is interposed between 2D output and 3D input during training allows to suppress this propagation of errors. As the 2D network was optimized during training on the training dataset, it is very rare that it wrongly marks a small scar in a single slice of a patient. However, on unseen data this is the case more often, especially if there are images of healthy patients in the cohort. For example on the EMIDEC dataset 2D nnU-Net marks some false positive scars in 20 out of the 33 healthy patients, which also mainly explains the big performance gap between 2D and 3D nnU-Net on this dataset. Although the vanilla cascade is able to correct these errors on 8 of the 20 patients, our method is able to correct the false positive scars in 16 cases. The reason for this is that by adding incorrect scar annotations during training our network was trained to be more prone to such false positive scar annotations also on

Table 7

Ablation study on the MyoPS test dataset. A comparison study of Dice scores regarding segmentation of myocardial infarction when trying to improve accuracy using a cascaded framework.

| Target | Methods | 2D Net Ours | 2D nnU-Net | 3D nnU-Net | Zhou et al. (2021a) | Zhang (2021) | Rahman and Marculescu (2023) |
|---|---|---|---|---|---|---|---|
| Infarction | Original | 63.24 ± 22.59* | 62.95 ± 24.89 | 64.20 ± 22.08 | 55.11 ± 27.21* | 62.04 ± 25.34* | 60.20 ± 28.18* |
| | Vanilla-cascade | 64.60 ± 22.85 | 62.92 ± 25.42* | 63.91 ± 21.80 | 56.54 ± 27.11 | 62.49 ± 25.41* | 61.86 ± 28.47 |
| | Ours | **66.21 ± 21.33** | **64.15 ± 23.81** | **65.02 ± 21.84** | **58.74 ± 24.59** | **63.78 ± 24.27** | **62.87 ± 27.35** |

* Indicates the method obtained statistically poorer results ($p < 0.05$) compared to the best performance, which is marked in bold.

unseen data. This is illustrated in Fig. 5, where our framework is able to detect an incorrectly marked scar on a healthy patient and removes it for the final segmentation prediction. This may also be the reason why our proposed method is somewhat less effective on the MyoPS dataset, as it does not contain images of healthy patients. A similar argument holds for cases where scars were not fully detected in all slices by the 2D framework. As a trained 2D network is very rarely missing scars on the training dataset it cannot be expected that such errors can get picked up easily by a cascaded framework at the testing stage. However, by introducing random erasure of scars in some slices we train our framework to also be more sensitive towards potentially missing scar pieces during inference. An example for this is illustrated in Fig. 4, were our cascaded method is able to fully detect a scar that was only partially marked by the 2D method.

In our ablation studies we showed that our method is able to improve segmentation accuracy for myocardial infarction, when given the segmentation results of various reference methods. For myocardial infarction the method trained with the proposed perturbation module was able to outperform the vanilla cascade pipeline on both datasets, see Tables 6 and 7. Further, the results of the ablation experiments provide an indication that our method works specifically well for 2D segmentation errors, as the improvements compared to the vanilla cascade framework turned out to be significant more frequently when applied to two dimensional models (Fig. 6). For example, on both datasets our method performed significantly better than the vanilla cascade framework when refining the outputs of 2D nnU-Net. On the other hand, when using the predictions of 3D nnU-Net no significant difference could be found between the two cascaded methods.

As we were able to significantly improve scar segmentation compared to state-of-the-art methods, the same effect could be expected for MVO segmentation. However, regarding MVO the proposed method performs very similar to the reference methods. When comparing the Dice coefficients for MVO segmentation of the different methods no significant differences were found. We suspect that this is the case because MVO is typically a much smaller structure that is not always visible across multiple layers. In the dataset we used there e.g. are examinations in which MVO only appears on one single slice in the whole CMR volume. This leads us to believe that the perturbations we introduced during training have a much smaller effect or no effect at all for the MVO class. We conjecture that the reason for this is that MVO occurs very rarely and does not always extend over several slices, which makes our approach less effective. However, the clinical information that MVO always occurs only within the center of a myocardial infarction could also be exploited for MVO segmentation. Therefore, including shape priors into the framework, for example using neural networks for shape reconstruction, as proposed by Yue et al. (2019), might be of interest for future work. Another possibility to improve the MVO segmentation could also be to incorporate classification priors to the framework as done by Brahim et al. (2022). By adding a binary classification module to the bottleneck of their segmentation network they were able to improve their methods accuracy in identifying MVO regions. Therefore, we believe that our method could also benefit from adding such classification priors to 2D or 3D frameworks, which may also be of interest for future work. Another approach to further improve detection of MVO in the future could be to also include clinical information about the patients into the segmentation framework.

The automatic segmentation tool developed in this study not only improves the accuracy of infarct size assessment, but could potentially be applied in clinical practice. Since the applied CNNs are not very large, we were able to compute clinical markers such as infarct size and MVO size on a conventional clinical computer without any GPU support within 3 to 5 s per patient. This could save a lot of time for radiologists, who have reported taking up to 20 min to manually quantify infarcts. By providing fast and accurate measurements of infarct severity and location this technology can improve patient outcomes by enhancing prognostication, supporting personalized treatment plans, and reducing subjectivity in infarct size assessment. We also believe that the cascaded structure of our framework could lead to benefits in clinical practice. For example, we could alert physicians to possible outliers or uncertainties when 2D and 3D segmentations show large differences.

Our thorough analysis of the impact of the different parts of the proposed pipeline to the segmentation performance using two challenging datasets with partially noisy images including various artifacts suggests that our framework may also be generally applicable to different applications. In our opinion, our framework can specifically be applied to three-dimensional segmentation problems with anisotropic pixel spacing where there is a limited number of manually segmented 3D volumes available. This could include tasks such as brain and prostate tumor segmentation in MR or liver lesion segmentation in CT images. Therefore, in future work, we plan to extend our method to several other challenging medical image segmentation tasks. We plan to investigate which types of 2D perturbations are effective for different segmentation problems, which will hopefully provide more insight into the developed training strategy.

A major limitation of our framework is its three-stage nature. To be able to successfully apply our framework for a segmentation task, one needs to train both first a 2D and then a 3D network. Moreover, the operations used in the perturbation module may need to be adapted to the different tasks. As the introduced perturbations are not differentiable it is also not possible to train both the 2D and the 3D models from start to finish, making the method a bit cumbersome. Another limitation is that both datasets we used do not reflect clinical reality. For example, all the data contains only slices in which myocardium of the left ventricle is visible. Furthermore, in the EMIDEC dataset images were registered in such a way that the left ventricle is in the center of each image and the examinations of the MyoPS challenge only include patients which suffered from myocardial infarction. However, when dealing with data in a clinical setting this cannot be expected. Therefore, to make the method applicable to clinical data, some preprocessing steps would still be necessary. Finally, we also want to mention that we could not test our method on the test dataset of the EMIDEC challenge which contains 50 additional examinations, since the manual labels for these images are not publicly available. This is a further limitation of our work, as an additional evaluation of the results on a separate test dataset would be desirable.

## 5. Conclusion

In this paper we describe a deep learning-based cascaded framework for a fully automatized segmentation of myocardial tissue in LGE CMR images. With the introduced perturbation module between the two segmentation networks during training, our framework is able to detect and correct segmentation errors which happen during the first segmentation step to a larger extent than previously proposed cascaded frameworks. Regarding segmentation of myocardial scars our methods was able to outperform various state-of-the-art reference methods on two different datasets. Further, in ablation studies we demonstrated the positive effects of the newly introduced perturbation module for improving the accuracy of infarction segmentation. Additionally our method proved to show myocardium, scar and MVO segmentations that are comparable to the manual segmentations. However, especially in MVO segmentation there is some room for improvement as not all cases of MVO were identified correctly by our method. Thus further studies which integrate prior information such as shape or also clinical meta-data information are desirable for future research to further improve the automatic detection of myocardial abnormalities.

**CRediT authorship contribution statement**

**Matthias Schwab:** Writing – original draft, Software, Methodology, Conceptualization. **Mathias Pamminger:** Writing – review & editing, Investigation. **Christian Kremser:** Writing – review & editing, Conceptualization. **Daniel Obmann:** Software. **Markus Haltmeier:** Writing – review & editing, Supervision, Conceptualization. **Agnes Mayr:** Writing – review & editing, Supervision, Project administration.

**Declaration of competing interest**

The authors declare that they have no known competing financial interests or personal relationships that could have appeared to influence the work reported in this paper.

**Acknowledgments**

This work was supported by the Austrian Science Fund (FWF) [grant number DOC 110].

**Appendix A. Supplementary data**

Supplementary material related to this article can be found online at https://doi.org/10.1016/j.media.2025.103594.

**Data availability**

Data will be made available on request.